\documentclass{article}
\usepackage{amsmath}
\usepackage{amssymb}
\usepackage{mathtools}

\usepackage{amsthm}

\usepackage{amsfonts}

\usepackage{stackengine}

\usepackage{biblatex}
\DeclareMathOperator{\Tr}{Tr}

\DeclarePairedDelimiterX{\inp}[2]{\langle}{\rangle}{#1, #2}
\title{MIMO Ambiguity Functions and Harmonic Analysis}
\author{Eren Berk Kama \and Mustafa Kuzuoğlu}
\date{}
\begin{document}
\maketitle
\begin{abstract}
Multi input multi output (MIMO) systems' capability of using seperate signals brings many advantages to radar signal processing and time frequency analysis. In this paper, a variety of properties of MIMO ambiguity functions related with representations of Heisenberg group are given. Some of the existing results for SIMO ambiguity functions are generalized to MIMO case. Combined effect of seperate signals is investigated.

\end{abstract}

\section{Introduction}

Ambiguity functions are useful signal processing tools, used in radar waveform design and time frequency signal processing. They were first given in \cite{woodward2014probability} In radar, they are used to analyze signal's behavior on the delay Doppler $(\tau,\nu)$ domain. In time frequency analysis, as they are related to Wigner distributions via a 2D-type Fourier transform (one Fourier transform and one inverse Fourier transform), they are used as the filter domain. More on use of ambiguity functions in time frequency analysis can be found in \cite{boashash2015time},\cite{grochenig2001foundations}.  As they are used frequently in these areas, there is a need to investigate their properties. Harmonic analysis studies the relations between a function and its Fourier transform counterpart, a good reference giving both commutative and noncommutative harmonic analysis is  \cite{folland2016course}.  Harmonic analysis methods are frequently used to analyze time frequency distributions. SIMO ambiguity functions were tied with harmonic analysis by using irreducible representations of Heisenberg group in \cite{auslander1985radar}, \cite{miller2002topics},\cite{moran2001mathematics}. Some applications of representations of Heisenberg group in signal processing are  \cite{auslander1995radar},\cite{schempp1984radar},\cite{auslander1979computing},\cite{howard2006finite},\cite{tolimieri1997time},\cite{auslander1993sliding}. More on Heisenberg group and harmonic analysis can be found in \cite{folland1989harmonic},\cite{howe1980role}. 
\newline
Ambiguity functions were generalized to MIMO in \cite{san2007mimo}. As MIMO systems are capable of using separable, incoherent signals, performance in detection, estimation and beampattern design is better compared with SIMO systems. In \cite{chen2008mimo}, a variety of MIMO ambiguity functions were given. Here, we will use representations of Heisenberg group to investigate properties of MIMO ambiguity functions. We will use some results in \cite{auslander1985radar}, \cite{miller2002topics},\cite{moran2001mathematics} in MIMO setting. As MIMO systems use multiple signals together, resulting ambiguity function consists of multiple self and cross ambiguity functions. Therefore, the combined effect of input signals is seen in the results. This brings new properties in terms of harmonic analysis, together with some constraints on the operations we can perform. 
\newline 
The rest of the paper is as follows. Section 2 gives a short preliminary on representation theory of Heisenberg group and definition of MIMO ambiguity functions. Section 3 gives properties of MIMO ambiguity functions. Section 4 gives some symmetry properties. Section 5 concludes the paper.

\section{Preliminaries}

\subsection{Representations of $H_\mathbb{R}$}
In this section, we will give a review of representation theory without going into too much detail. Ideas we present here, their proofs and more on representations of Heisenberg group can be found in harmonic analysis books which cover noncommutative harmonic analysis topics, a good reference is \cite{folland2016course}. Especially representations of Heisenberg group in relation with ambiguity functions can be found in \cite{auslander1985radar}, \cite{miller2002topics},\cite{moran2001mathematics}. In \cite{miller2002topics}, ideas are constructed starting from group theory and the relationship with wavelets is given (via representation of affine group ).
\newline
A representation of a group $G$, with its representation space $V$ is a homomorphism $T : g \mapsto T(g)$ of $G$ into $GL(V)$. $T : G \rightarrow GL(V)$. $GL(V)$ is the set of all linear transformations from $V$ to $V$.  We will use representations as operators here to keep familiarity with operations in signal processing applications. Matrix representation could be obtained by use of an orthonormal basis. We will be mainly working on $L_2(\mathbb{R})$, which is the space of signals with finite energy. A unitary operator on $L_2(\mathbb{R})$ is a linear mapping satisfying
\newline
 \begin{equation*}
  \inp*{T\mathbf v}{T\mathbf w} =
  \inp{\mathbf v}{\mathbf w} \qquad
  \forall \mathbf v, \mathbf w \in V
\end{equation*}
 A very important class of representations is irreducible representations. A representation is reducible if there exists a subspace $W$ of $V$ such that $T(g)w \in W$ for every $g \in G$, $w \in W$. Otherwise it is irreducible. There are different constructions of Heisenberg groups, here we will use the following,
\newline
Heisenberg group is a group of matrices
 \begin{equation*}
H_\mathbb{R} = \{(x_1,x_2,x_3)= \begin{bmatrix} 
1 & x_1 & x_3  \\
0 & 1 & x_2  \\
0 & 0 & 1
\end{bmatrix} : x_i \in \mathbb{R}           	\}
\end{equation*}
with group product
 \begin{equation*}
(x_{1},x_2,x_3)(y_1,y_2,y_3)=(x_1+y_1,x_2+y_2,x_3+y_3+x_1y_2)
 \end{equation*}
Representations of Heisenberg group can be found by induction. Construction of representations can be found in \cite{folland2016course}. Here, we will not go into details of constructing representations, rather, we will use them. We will use the following relation in writing ambiguity functions. The Schroedinger representation of $H_\mathbb{R}$ is, 
 \begin{equation*}
T(x_1,x_2,x_3)f(t) = e^{i2\pi (x_{3}+x_{2}t)}f(t+x_{1})
\end{equation*}
T is an irreducible representation of $H_\mathbb{R}$ on $L_2(\mathbb{R})$. In ambiguity functions, we wont be using the element ${x_3}$, we will use $T(x_1,x_2,0)$. We will construct ambiguity functions from irreducible representations on the next subsection. Now, we will mention some properties of representations. Representations are closely related with positive definite functions. A positive definite function is a function which satisfies, 
 \begin{equation*}
\sum_{ij=1}^{N} \rho(x_{j}^{-1} x_{i}) \bar{c_{j}} c_{i} \geq 0
\end{equation*}
for any set of complex numbers $c_{1},c_{2},...,c_{N}$, and elements $x_{i}$ of the group $G$. Let $T$ be a unitary representation of G on $L_2(\mathbb{R})$ (or any Hilbert space $\mathcal{H}$). Positive definite functons can be written via representations, 
\begin{equation*}
  p = \inp*{T\mathbf v}{\mathbf v} 
\end{equation*}
These relations will be used on ambiguity functions in the next section.


If $T$ is an irreducible representation, then $\{T(x)\mathbf u\}$ spans a dense subspace of $L_2(\mathbb{R})$.
\newline
An important property of irreducible representations is that, positive definite functions created by an irreducible representation are indecomposable. A positive definite function $p$ is indecomposable if any other positive definite function $p'$, where $p-p'$ is also positive definite, is a scalar multiple of $p$.

\subsection{MIMO Ambiguity Functions}
We will give the definition of ambiguity functions here, more on radar signal processing can be found in \cite{richards2014fundamentals},\cite{levanon2004radar}. We may write the traditional self ambiguity functions as follows \cite{levanon2004radar}
\begin{equation*}
 \chi (\tau,\nu)  =  \int_{-\infty}^{+\infty} u(t) u^*(t + \tau) e^{j2\pi\nu} dt 
\end{equation*}
$u(t)$ being the signal used as input. This function shows the matched filter output where there is a delay mismatch $\tau$ and Doppler mismatch $\nu$. In order to achieve high resolution in both range and Doppler, one desires to make the function  $\chi (\tau,\nu)$ sharp around $(0,0)$ point.
The cross ambiguity function is defined as follows,
\begin{equation*}
 \chi (u,v) (\tau,\nu)  =  \int_{-\infty}^{+\infty} u(t) v^*(t + \tau) e^{j2\pi\nu} dt 
\end{equation*}
A very closely related time frequency distribution is the Wigner distribution which is defined as,
\begin{equation*} 
W (u,v)(t,f)  =  \int_{-\infty}^{+\infty} u(t+ \tau/2) v(t - \tau/2) e^{-j2\pi\nu t} d\tau
\end{equation*}
We can write the ambiguity function in terms of representations of $H_\mathbb{R}$ as,
 \begin{equation*}
 \inp {\mathbf u}{T({\tau},{\nu})\mathbf v} = \int_{-\infty}^{+\infty}  u(t) v^*(t+\tau) e^{i2\pi \nu t}  dt
\end{equation*}
Here, we used the operation on the second signal for notational convention. It gives conjugate of the Doppler shift when we use the first representation we presented, though we may assume here as the $x_{2}$ axis is inverted. This doesn't change any result, it merely is done for notational convention.
MIMO ambiguity functions were given by \cite{san2007mimo}. Here we will review the definiton. As we have mentioned in the introduction, \cite{chen2008mimo} gave first properties in terms of waveform design. We will use the notation of \cite{chen2008mimo}, \cite{levanon2004radar}. Assumptions on ambiguity functions change the form of the function such as in narrowband and wideband radars. We will assume that sensors are close, such that target's velocity vector has similar effect in each sensor. In addition, target is in the farfield and bandwith is narrow to allow narrowband case. Together with these assumptions, in \cite{san2007mimo}, it was shown that  elements of the MIMO covariance matrix are the Woodward ambiguity functions depending only on time delay and Doppler shift.
We can write the MIMO correlation matrix as,
 \begin{equation*}
  \mathbf R_{ij}(\tau, \nu) = A_{ij}(\tau, \nu) = \int_{-\infty}^{+\infty} u_{i}(t) u_{j}^*(t - \tau) e^{j2\pi\nu} dt 
\end{equation*}
Elements $\mathbf R_{ij}(\tau, \nu)$ are the Woodward ambiguity functions. The MIMO ambiguity function is obtained by using the MIMO correlation matrix with the corresponding steering vectors.  
 \begin{equation*}
\chi(\tau,\nu, f_s, f^{'}_s) = \sum^{M-1}_{m = 0 } \sum^{M-1}_{{m^{'}} = 0 } \chi_{m,m^{'}}(\tau,\nu)  e^{i2\pi \gamma (f_s m - f^{'}_s m^{'} )} 
\end{equation*}
where $\chi_{m,m^{'}}(\tau,\nu) = \inp { u_m}{T({\tau},{\nu}) u_{m^{'}}}$ is the cross ambiguity function. As we see, in addition to delay and Doppler, in MIMO there is the normalized spatial frequency of the target $f_s$.  If there is no mismatch, the MIMO ambiguity function is $\chi(0,0, f_s, f_s)$. Therefore, for high resolution, we would like $\chi(\tau,\nu, f_s, f^{'}_s)$ to be sharp around the point $(0,0, f_s, f_s)$. Here, we note instead of discarding $x_3$ term in $T$, the representation could be written with $x_3=\gamma (f_s m - f^{'}_s m^{'} )$ to give the relation $e^{i2\pi \gamma (f_s m - f^{'}_s m^{'} )}$ with the use of representation. If we denote the new representation $T^{'}$ we have,
 \begin{equation*}
\chi(\tau,\nu, f_s, f^{'}_s) = \sum^{M-1}_{m = 0 } \sum^{M-1}_{m^{'} = 0 } \inp { u_m}{T^{'}({\tau},{\nu},x) u_{m^{'}}}
\end{equation*}
But using the $x_3$ term violates square integrability of ambiguity functions. Therefore, it is not very useful for our purposes, and we will not be using it.

\subsection{The Special Linear Group $SL(2,\mathbb{R})$}
In \cite{auslander1985radar}, action of the special linear group on SIMO ambiguity functions was given. In section 4, we will use this relations in MIMO setting and relate to signals of interest in time frequency analysis. $SL(2,\mathbb{R})$  is the group of $2\times 2$ real matrices, with determinant 1. It is called the special linear group. 
\begin{equation*}
SL(2,\mathbb{R})=\{ \begin{bmatrix} 
a & b  \\ 
c & d 
\end{bmatrix}
, \quad a,b,c,d \in \mathbb{R} \}
\end{equation*}
We will look at the following generators of $SL(2,\mathbb{R})$,

\begin{equation*}
J = \begin{bmatrix} 
0 & 1  \\ 
-1 & 0 
\end{bmatrix}
\end{equation*}

\begin{equation*}
t(a)= \begin{bmatrix} 
1 & 0  \\ 
a & 1 
\end{bmatrix}
, \quad a \in \mathbb{R} 
\end{equation*}

\begin{equation*}
m(b)= \begin{bmatrix} 
b & 0  \\ 
0 & 1/b 
\end{bmatrix}
, \quad b>0 
\end{equation*}

\section{Properties of MIMO Ambiguity Functions}
We will relate MIMO ambiguity functions with irreducible unitary representations of Heisenberg group and give a varierty of properties in various situations. We will use some results from \cite{auslander1985radar},\cite{miller2002topics},\cite{moran2001mathematics} in MIMO ambiguity functions. We note that relations in these references were given for SIMO case. Therefore, they were about self ambiguity of a single signal or cross ambiguity of two signals. We are using M self ambiguity functions and $M(M-1)$ cross ambiguity functions, and we are investigating in their combined effect. Throughout the paper, properties without prime are for traditional ambiguity functions and properties with prime on them are for MIMO ambiguity functions.

\begin{description}
\item[Property 3.1]
\end{description}
For $\chi(u_{m},u_{m^{'}}) \in L_2(\mathbb{R}^{2}) \; $ and $\; u_{m},u_{m^{'}} \in L_2(\mathbb{R})$

\begin{equation*} 
\Vert \chi_{m,m^{'}}(\tau,\nu) \Vert_{2}^{2} =  \Vert u_{m} \Vert^{2}\Vert u_{m^{'}} \Vert^{2}
\end{equation*}

\begin{proof} 
Fix $\tau$,
 \begin{equation*}
\Vert \chi_{m,m^{'}}(\tau,\nu) \Vert_{2}^{2} = \int \int_{-\infty}^{+\infty}   \vert \mathcal{F} \{ u_{m}(t)u^{*}_{m'}(t+\tau) \} \vert d\nu d\tau 
\end{equation*}
By Plancherel theorem (sometimes called generalized Parseval theorem in signal processing books),
 \begin{equation*}
= \int \int_{-\infty}^{+\infty}   \vert u_{m}(t)u_{m^{'}}(t+\tau) \vert^{2} dt d\tau = \Vert u_{m} \Vert^{2}\Vert u_{m^{'}} \Vert^{2}
\end{equation*}
\end{proof} 
This property means that, norm of $\chi_{m,m^{'}}(\tau,\nu) $ is equal to waveforms' energies and it is a constant, only depends on waveforms' energy. The following property takes this property to the MIMO case. A very similar property for unit energy signals was given in \cite{chen2008mimo}, 

\begin{description}
\item[Property 3.1']
\end{description}
We will assume $\gamma$ is an integer as in \cite{chen2008mimo},

\begin{equation*} 
\int_{0}^{1} \int_{0}^{1} \int \int_{-\infty}^{+\infty}   \vert \chi(\tau,\nu, f_s, f^{'}_s) \vert^{2}  d\tau d\nu df_s df^{'}_s = \sum^{M-1}_{m = 0 } \sum^{M-1}_{m^{'} = 0 } \Vert u_{m} \Vert^{2}\Vert u_{m^{'}} \Vert^{2}
\end{equation*}

\begin{proof} 
Again we will use Parseval's relation

 \begin{equation*}
\int_{0}^{1} \int_{0}^{1} \int \int_{-\infty}^{+\infty}   \vert \chi(\tau,\nu, f_s, f^{'}_s) \vert^{2}  d\tau d\nu df_s df^{'}_s = \\
 \end{equation*}
 \begin{equation*}
  \begin{aligned}
  \int \int_{-\infty}^{+\infty}\int_{0}^{1} \int_{0}^{1}   \vert \sum^{M-1}_{m = 0 } \sum^{M-1}_{{m^{'}} = 0 } \chi_{m,m^{'}}(\tau,\nu)  e^{i2\pi \gamma (f_s m - f^{'}_s m^{'} )}  \vert^{2}   df_s df^{'}_s d\tau d\nu
\end{aligned}
\end{equation*}
By a change of variables,
 \begin{equation*}
  \begin{aligned}
{1/\gamma^{2}} \int \int_{-\infty}^{+\infty}  \int_{0}^{\gamma} \int_{0}^{\gamma}  \vert \sum^{M-1}_{m = 0 } \sum^{M-1}_{{m^{'}} = 0 } \chi_{m,m^{'}}(\tau,\nu)  e^{i2\pi  (f_s m - f^{'}_s m^{'} )}  \vert^{2}   df_s df^{'}_s d\tau d\nu
\end{aligned}
\end{equation*}
 \begin{equation*}
= \int \int_{-\infty}^{+\infty}  \sum^{M-1}_{m = 0 } \sum^{M-1}_{{m^{'}} = 0 } \vert \chi_{m,m^{'}}(\tau,\nu) \vert^{2} dt d\tau = 
\end{equation*}
 \begin{equation*}
 = \sum^{M-1}_{m = 0 } \sum^{M-1}_{{m^{'}} = 0 } \Vert \chi_{m,m^{'}}(\tau,\nu) \Vert^{2} = \sum^{M-1}_{m = 0 } \sum^{M-1}_{{m^{'}} = 0 } \Vert u_{m} \Vert^{2}\Vert u_{m^{'}} \Vert^{2}
\end{equation*}
\end{proof} 
Here, we see that the energy of MIMO ambiguity is directly determined by energies of input signals. The following property gives the inner product relation between two ambiguity functions.

\begin{description}
\item[Property 3.2]
\end{description}
Let $u_{1},u_{2},u_{3},u_{4} \in L_2(\mathbb{R})$, then
\begin{equation*} 
\inp{\chi(u_{1},u_{3})}{\chi(u_{2},u_{4})}_{L_2(\mathbb{R}^{2})}  = \inp{u_{1}}{u_{2}} \inp{u_{3}}{u_{4}} 
\end{equation*}

\begin{proof} 
If $\chi(u_{i},u_{j}) \in L_2(\mathbb{R})$ for $i=1,2,3,4$

 \begin{equation*}
 \int \int_{-\infty}^{+\infty}   \chi(u_{1},u_{3})\chi(u_{2},u_{4})  d\tau d\nu  
 \end{equation*}
 \begin{equation*}
 = \int \int_{-\infty}^{+\infty}    \mathcal{F} \{ u_{1}(t)u_{3}^*(t+\tau) \} \mathcal{F} \{ u_{2}(t)u_{4}^*(t+\tau) \} d\nu d\tau 
 \end{equation*}
Again, by similar arguments to property 3.1,
 \begin{equation*}
  \begin{aligned}
= \inp{u_{1}}{u_{2}} \inp{u_{3}}{u_{4}} 
\end{aligned}
\end{equation*}
\end{proof}
This is a frequently used identity. Using this in definition of ambiguity functions,

 \begin{equation*}
\int \int_{-\infty}^{+\infty} \inp{u_{1}}{Tu_{3}} \inp{u_{2}}{Tu_{4}}   d\tau d\nu = \inp{u_{1}}{u_{2}} \inp{u_{3}}{u_{4}} 
\end{equation*}
Miller \cite{miller2002topics} stated that, if we use an orthonormal basis $\{u_{i}\}$ of $L_2(\mathbb{R})$ in constructing ambiguity functions, they form an orthonormal basis for $L_2(\mathbb{R}^{2})$. In order for ambiguity functions to be an orthonormal basis of $L_2(\mathbb{R}^{2})$, they need to be dense in $L_2(\mathbb{R}^{2})$. To show that ambiguity functions are dense in $L_2(\mathbb{R}^{2})$,  it is enough to show that $\inp{\chi(u_{i},u_{j})(\tau,\nu)}{h(\tau,\nu)}_{L_2(\mathbb{R}^{2})}=0$ for all $i,j$ implies $h(\tau,\nu)=0$ almost everywhere. A simple proof of this can be found in \cite{auslander1985radar}. The normality part here is obtained by using signals with unit norm. We may drop the normality here. We note that by similar arguments Wigner distributions can be shown to hold the same relation. This relation allows us to write signals in $L_2(\mathbb{R}^{2})$ with basis expansions. This can be used in time frequency detection, estimation of LTV or nonstationary systems and filtering applications. 
\newline
As we use a combination of different signals in MIMO ambiguity functions, we are actually using $M^{2}$ elements of the orthogonal basis of $L_2(\mathbb{R}^{2})$. This relation, together with MIMO Wigner distributions can be used in time frequency signal processing problems.

%

\begin{description}
\item[Property 3.2']
\end{description}
Let $u_{i},v_{j} \in L_2(\mathbb{R})$ for all ${i},{j}$ and let $\chi_{u}(\tau,\nu, f_s, f^{'}_s)$ and $\chi_{v}(\tau,\nu, f_s, f^{'}_s)$ denote the ambiguity function from waveforms $u_{i}$ and $v_{j}$ respectively, then

\begin{equation*} 
\inp{\chi_{u}(\tau,\nu, f_s, f^{'}_s)}{\chi_{v}(\tau,\nu, f_s, f^{'}_s)} =
\end{equation*}

\begin{equation*} 
 = \sum_{m,m^{'} } \sum_{n,n^{'} } \inp{u_{m}}{v_{n}}    \inp{v_{n^{'}}}{u_{m^{'}}} e^{i2\pi \gamma [(f_s m - f^{'}_s m^{'} )-(f_s n - f^{'}_s n^{'})]}
\end{equation*}

\begin{proof} 
Proof can be done by inserting property 3.2 in this relation

 \begin{equation*}
\inp{\chi_{u}(\tau,\nu, f_s, f^{'}_s)}{\chi_{v}(\tau,\nu, f_s, f^{'}_s)} = 
 \end{equation*}
 \begin{equation*}
 = \left( \sum_{m,m^{'} } \chi_{m,m^{'}}(\tau,\nu)  e^{i2\pi \gamma (f_s m - f^{'}_s m^{'} )} \right)  \left(\sum_{n,n^{'} } \chi_{n,n^{'}}(\tau,\nu)  e^{i2\pi \gamma (f_s n - f^{'}_s n^{'} )} \right)^* \\
 \end{equation*}
 \begin{equation*}
  \begin{aligned}
= \sum_{m,m^{'} }\sum_{n,n^{'} } \chi_{m,m^{'}}(\tau,\nu) \chi_{n,n^{'}}(\tau,\nu) e^{i2\pi \gamma (f_s m - f^{'}_s m^{'} )} e^{i2\pi \gamma (f_s n - f^{'}_s n^{'} )}
\end{aligned}
\end{equation*}

\begin{equation*}
  \begin{aligned}
= \sum_{m,m^{'} } \sum_{n,n^{'} } \inp{u_{m}}{v_{n}}    \inp{v_{n^{'}}}{u_{m^{'}}} e^{i2\pi \gamma [(f_s m - f^{'}_s m^{'} )-(f_s n - f^{'}_s n^{'})]}
\end{aligned}
\end{equation*}

\end{proof} 
Here, we again see that if $\{ u_{i} \}\{v_{j} \}$ together form an orthonormal basis for $L_2(\mathbb{R})$, and if we fix $f_s, f_s^{'}$, then $\chi_{u}(\tau,\nu, f_s, f^{'}_s)$ is an orthonormal basis in $L_2(\mathbb{R}^{2})$. That they are dense in $L_2(\mathbb{R}^{2})$ follows from the same argument for property 3.2. 
We may look at the following case of two MIMO ambiguity functions created by using all orthonormal signals.

\begin{description}
\item[Property 3.2'']
\end{description}
 If $\{ u_{i} \}\{v_{j} \}$ together is an orthonormal basis of $L_2(\mathbb{R})$, then the inner product of two MIMO ambiguity functions created with these signals is,

\begin{equation*} 
\sum_{m,m^{'} } \sum_{n,n^{'} } \inp{u_{m}}{v_{n}}    \inp{v_{n^{'}}}{u_{m^{'}}} e^{i2\pi \gamma [(f_s m - f^{'}_s m^{'} )-(f_s n - f^{'}_s n^{'})]} = M^{2}
\end{equation*}

\begin{proof} 
If we use the orthogonality of $u_{i},v_{j}$,

 \begin{equation*}
\inp{u_{m}}{v_{n}}  \inp{v_{n^{'}}}{u_{m^{'}}} =  \delta_{mn} \delta_{{m^{'}}{n^{'}}}
 \end{equation*}

 \begin{equation*}
  \begin{aligned}
 \sum_{m,m^{'} } \sum_{n,n^{'} } \delta_{mn} \delta_{{m^{'}}{n^{'}}} e^{i2\pi \gamma [(f_s m - f^{'}_s m^{'} )-(f_s n - f^{'}_s n^{'})]} = \sum_{m,m^{'} } \sum_{n,n^{'} }  e^{i0} = M^{2}
\end{aligned}
\end{equation*}
\end{proof} 

Properties 3.3, 3.4 and 3.5 are given in \cite{auslander1985radar},\cite{miller2002topics},\cite{moran2001mathematics}, we will repeat them here and use them in MIMO case.
The following property uses positive definiteness relation of representations mentioned in section 2 in ambiguity functions.

 \begin{description}
\item[Property 3.3]
\end{description}
$\chi(\tau,\nu)$ is a positive definite function.

\begin{proof} 
$T$ is a unitary representation of the Heisenberg group and $\chi(u,v)(\tau,\nu) = \inp{T(x)u}{u}$. Therefore, from the definition in Section 2, $\chi(\tau,\nu)$ is a positive definite function.
\end{proof}
From this property, we see that self ambiguity functions are positive definite as a result of representations. An intereseting point we can investigate about MIMO ambiguity functions is the no spatial frequency mismatch part of the function's support, which is the part corresponding to $(\tau,\nu,f_s, f_s)$. As  there are cross terms in MIMO ($u_{m} \neq u_{m'}$), rather than self terms, we are interested in finding cases where we can find self terms or cases where $u_{m} = u_{m'}$. In the next property, we will look at self terms by taking the integral on the no spatial frequency mismatch part similar to \cite{chen2008mimo}. But we will still have the delay Doppler variables, we wont look at the identity point.

 \begin{description}
\item[Property 3.3']
\end{description}
If $\gamma$ is an integer $\int_{0}^{1} \chi(\tau,\nu,f_s, f_s) df_{s}$ is a positive definite function

\begin{proof} 
 \begin{equation*}
\int_{0}^{1} \sum^{M-1}_{m = 0 } \sum^{M-1}_{{m^{'}} = 0 }  \chi_{m,m^{'}}(\tau,\nu) e^{i2\pi \gamma (f_s m - f_s m^{'} )} df_{s} = 
 \end{equation*}

 \begin{equation*}
  \begin{aligned}
=  \sum^{M-1}_{m = 0 } \sum^{M-1}_{{m^{'}} = 0 }  \chi_{m,m^{'}}(\tau,\nu) \delta_{m,m^{'}} = \sum^{M-1}_{m = 0 }  \chi_{m,m}(\tau,\nu)
\end{aligned}
\end{equation*}
Last equation follows as delta will only leave the self terms. By this integration on no spatial frequency mismatch part, self ambiguity terms are obtained. In Property 3.3, we have shown that self ambiguity functions $\chi_{m,m}$ are positive definite. Therefore, $\sum^{M-1}_{m = 0 }  \chi_{m,m}(\tau,\nu)$ is also a positive definite function. Here we see that $\int_{0}^{1} \chi(\tau,\nu,f_s, f_s) df_{s}$ is positive definite.
\end{proof}
Here, we note that the relation $\sum^{M-1}_{m = 0 }  \chi_{m,m}(\tau,\nu)$ is equal to the trace of the MIMO correlation matrix $\Tr (\mathbf R(\tau,\nu))$. Therefore, the result holds for the trace of the correlation matrix too. Next two properties use positive definiteness to give a constraint on signals creating the same ambiguity function. We will give a proof similar to that in \cite{auslander1985radar} for property 3.4 as it is simpler. A different proof, with more comment on the algebraic structure can be found in \cite{miller2002topics}, we will not be using the proof, we give it here for completeness.

 \begin{description}
\item[Property 3.4]
\end{description}
If $u,v \in L_2(\mathbb{R})$ and $\chi(u,u) = \chi(v,v)$, then

 \begin{equation*}
u= \lambda v
 \end{equation*}
for a constant $\lambda \in \mathbb{C}$., with $\vert \lambda \vert = 1$

\begin{proof} 
As given in section 2, the span of $\{ T_{x}u\}$ is dense in $L_2(\mathbb{R}^{2})$ for an irreducible $T$. Let $x,y,z \in H_\mathbb{R}$, define $S$ such that, $ S(T_{x}u) = T_{x}v $. We have $\inp{T_{x}u}{u} = \inp{T_{x}v}{v}$ then, $\inp{T_{y}u}{T_{x}u} = \inp{T_{y}v}{T_{x}v}$ holds for all $x,y \in H_\mathbb{R}$. Suppose, $T_{y}u=T_{z}u$, $\inp{T_{y}v}{T_{x}v} = \inp{T_{y}v}{T_{x}v}$ for all $x \in H_{\mathbb{R}}$

As $\{T_{x}v\}$ spans a dense subspace of $L_2(\mathbb{R})$, $T_{y}v = T_{x}v$ also holds. Therefore, $S: \{T_{x}u \} \rightarrow \{ T_{x}v\}$ is a unitary operator of $L_2(\mathbb{R})$ satisfying $ST(x) = T(x)S $ for all $x \in H_{\mathbb{R}}$. As a result, we have $ST(x)S^{-1} = T(x) $ and $S=\lambda I $, $\vert \lambda \vert = 1$, which implies the result we stated.
\end{proof}

 \begin{description}
\item[Property 3.5]
\end{description}
If $\chi(u_1,u_1) = \chi(u_2,u_2) + \chi(u_3,u_3)  $ for $u_1,u_2,u_3 \in L_2(\mathbb{R})$ then $u_2 =\alpha u_3$ for $\alpha \in \mathbb{C}$.

\begin{proof} 
Let $p_{i} = \chi(u_i,u_i) = \inp{u_i}{Tu_i}$, then we know that $p_{i}(x)$ is a positive definite function by the arguments in property 3.3. As $\chi(u_1,u_1) = \chi(u_2,u_2) + \chi(u_3,u_3)  $, we have $p_{1} = p_{2} + p_{3}$. Therefore, $p_{1}$ dominates both $p_{2}$ and $p_{3}$. As we mentioned in section 2, positive definite functions constructed by irreducible representations  are indecomposable, which implies that $p(x) = \inp{u}{Tu}$ is indecomposable. As a result, the positive definite functions are related as, $p_{1} = \beta_{2}^{2} p_{2} = \beta_{3}^{2}  p_{3}$, and $\inp{u_1}{Tu_1} = \inp{\beta_{2} u_2}{T \beta_{2} u_2} = \inp{\beta_{3} u_3}{T \beta_{3} u_3}$. Similar to property 3.4, we have that $\beta_{2} u_2 = \lambda \beta_{3} u_3 $, combining terms, we obtain $\alpha = \beta_{2}/\beta_{3} \lambda$. Therefore, sum of two ambiguity functions is again an ambiguity function iff $u_2 = \alpha u_3$
\end{proof}
In the next property, we will use property 3.5 on MIMO ambiguity functions.
 \begin{description}
\item[Property 3.5']
\end{description}
If $\gamma $ is an integer,
 \begin{equation*}
\int_{0}^{1} \chi(\tau,\nu,f_s, f_s) df_{s}
 \end{equation*}
Can be interpreted as an ambiguity function iff $u_{i} = \lambda_{ij} u_{j}$, $\vert \lambda_{ij} \vert = 1 $ for all $i,j = 0,1,...,M-1$

\begin{proof} 
Proof of this property uses the same arguments with that of property 3.5'
 \begin{equation*}
\int_{0}^{1} \sum^{M-1}_{m = 0 } \sum^{M-1}_{{m^{'}} = 0 }  \chi_{m,m^{'}}(\tau,\nu) e^{i2\pi \gamma (f_s m - f_s m^{'} )} df_{s} = 
 \end{equation*}

 \begin{equation*}
  \begin{aligned}
=  \sum^{M-1}_{m = 0 } \sum^{M-1}_{{m^{'}} = 0 }  \chi_{m,m^{'}}(\tau,\nu) \delta_{m,m^{'}} = \sum^{M-1}_{m = 0 }  \chi_{m,m}(\tau,\nu) 
\end{aligned}
\end{equation*}
We can denote $\sum^{M-1}_{m = 0 }  \chi_{m,m}(\tau,\nu) $ as $p_A$. We know that $\chi_{m,m}$ are positive definite and indecomposable and $p_A$ dominates all $\chi_{m,m}$. Therefore, $p_A$ is an ambiguity function iff $u_{i} = \lambda_{ij} u_{j}$, $\vert \lambda_{ij} \vert = 1 $ for all $i,j = 0,1,...,M-1$

\end{proof}
Again, we can see that the result holds for the trace of correlation matrix $\Tr (\mathbf R(\tau,\nu))$. This means that, the sum of self ambiguity functions can be treated as a single ambiguity function iff input signals satisfy $u_{i} = \lambda_{ij} u_{j}$, $\vert \lambda_{ij} \vert = 1 $ for all $i,j = 0,1,...,M-1$. This may be seen as a reduction in the number of terms by $M-1$ on the trace of correlation matrix.

\section{Symmetry Properties of MIMO Ambiguity \newline Functions}
Here, we will give the actions of the group $SL(2,(\mathbb{R}))$ on ambiguity functions. The actions can be realized as $\chi(u,v)(M(\tau,\nu)) = \chi(u,v)(\tau,\nu) \circ {M}$, where $M$ is an $2 \times 2$ invertible matrix with determinant $1$. We will observe which signal creates the signal after transform. This will help to realise the set of signals and corresponding ambiguity functions that can be obtained via the combined actions of tinvertible $2 \times 2$  matrices.

\begin{description}
\item[Property 4.1]
\end{description}
Let $u,v \in L_2(\mathbb{R}) $, then

\begin{equation*} 
\chi(u,v)(\tau,\nu) \circ {J} = \chi(\mathcal{F}\{u\},\mathcal{F}\{v\})(\tau,\nu) e^{j2\pi\nu \tau}
\end{equation*}

\begin{proof} 
We will use definition of ambiguity functions

 \begin{equation*}
\chi(u,v)(\tau,\nu) = e^{j2\pi\nu \tau} \inp{u(t)}{v(t +\tau)e^{j2\pi(t+ \tau)}}
 \end{equation*}
 \begin{equation*}
  \begin{aligned}
= e^{j2\pi\nu \tau} \inp{\mathcal{F}\{u\}(\xi)}{\mathcal{F}\{v\}(\xi-\nu) e^{j2\pi \tau (\xi-\nu)}}
\end{aligned}
\end{equation*}
 \begin{equation*}
  \begin{aligned}
=  e^{j2\pi\nu \tau} \chi(\mathcal{F}\{u\},\mathcal{F}\{v\})(-\nu,\tau)
\end{aligned}
\end{equation*}
Therefore, 
 \begin{equation*}
\chi(\mathcal{F}\{u\},\mathcal{F}\{v\})(\tau,\nu) = e^{-j2\pi\nu \tau} \chi(u,v)(\nu, -\tau)
\end{equation*}
 \begin{equation*}
  \begin{aligned}
=  e^{j2\pi\nu \tau} \chi(u,v)(\tau,\nu) \circ {J} 
\end{aligned}
\end{equation*}
The property follows
\end{proof} 
Next, we will use this in MIMO ambiguity functions,

\begin{description}
\item[Property 4.1']
\end{description}
If we fix $f_s,f'_s$ and denote $\chi_{u}(\tau,\nu, f_s, f'_s)$ the MIMO ambiguity function with the set of waveforms $\{ u_m\}$,

\begin{equation*} 
\chi_{u}(\tau,\nu, f_s, f'_s) \circ {J} = \chi_{\mathcal{F}\{u\}}(\tau,\nu, f_s, f'_s) e^{j2\pi\nu \tau} = \chi_{u}(\nu, -\tau, f_s, f'_s)
\end{equation*}

\begin{proof} 
We will use definition of ambiguity functions

 \begin{equation*}
\chi_{\mathcal{F}\{u\}}(\tau,\nu, f_s, f'_s) = \sum^{M-1}_{m = 0 } \sum^{M-1}_{{m^{'}} = 0 } \chi(\mathcal{F}\{u_m\},\mathcal{F}\{u_m^{'}\})(\tau,\nu)  e^{i2\pi \gamma (f_s m - f^{'}_s m^{'} )} 
 \end{equation*}
 \begin{equation*}
  \begin{aligned}
= \sum^{M-1}_{m = 0 } \sum^{M-1}_{{m^{'}} = 0 } \chi(u_m,u_m^{'})(\nu, -\tau)  e^{i2\pi \gamma (f_s m - f^{'}_s m^{'} )} 
\end{aligned}
\end{equation*}
 \begin{equation*}
  \begin{aligned}
= \sum^{M-1}_{m = 0 } \sum^{M-1}_{{m^{'}} = 0 } \chi(u_m,u_m^{'})(\tau,\nu) \circ J e^{-j2\pi\nu \tau}   e^{i2\pi \gamma (f_s m - f^{'}_s m^{'} )} 
\end{aligned}
\end{equation*}
 \begin{equation*}
= \chi_{u}(\tau,\nu, f_s, f'_s) \circ {J}  e^{-j2\pi\nu \tau} = \chi_{u}(\nu, -\tau, f_s, f'_s)
\end{equation*}
By moving the exponential to the other side of equations, the property follows.
\end{proof} 
This property shows that $J$ acts on MIMO ambiguity functions to give ambiguity function of the Fourier transforms of $\{ u_m\}$. next property will be twice application of $J$ on ambiguity functions.

\begin{description}
\item[Property 4.2]
\end{description}

\begin{equation*} 
(\chi(u,v)(\tau,\nu) \circ {J} ) \circ {J} = \chi(u,v)(-\tau,-\nu) = \chi^*(v,u)(\tau,\nu)
\end{equation*}

\begin{proof} 
Similar to property 4.1's proof, we will use the definition of ambiguity functions

 \begin{equation*}
\chi(u,v)(\tau,\nu) \circ {J} =\chi(u,v)(\nu, -\tau) e^{j2\pi\nu \tau} 
 \end{equation*}
 \begin{equation*}
  \begin{aligned}
(\chi(u,v)(\nu, -\tau) e^{j2\pi\nu \tau} ) \circ J = \chi(u,v)(-\tau,-\nu)
\end{aligned}
\end{equation*}
For the second equality, we can use a change of variables,
 \begin{equation*}
  \begin{aligned}
 \chi(u,v)(-\tau,-\nu) = \int_{-\infty}^{+\infty} u(t) v^*(t + \tau) e^{j2\pi\nu t} dt 
\end{aligned}
\end{equation*}
 \begin{equation*}
  \begin{aligned}
= \int_{-\infty}^{+\infty} u(t' + \tau) v^*(t' ) e^{j2\pi\nu (t'+\tau)} dt'
\end{aligned}
\end{equation*}
 \begin{equation*}
= \chi^*(v,u)(\tau,\nu) e^{-j2\pi\nu \tau}
\end{equation*}
\end{proof} 

Next property is application of this result to the MIMO case.
\newpage

\begin{description}
\item[Property 4.2']
\end{description}
If we fix $f_s,f'_s$ 

\begin{equation*} 
(\chi(\tau,\nu, f_s, f'_s) \circ {J}) \circ J =  \chi( -\tau, -\nu, f_s, f'_s) = \chi^*(\tau,\nu, f'_s, f_s) e^{-j2\pi\nu \tau}
\end{equation*}

\begin{proof} 
For both equations, we can use property 4.2 in definition of MIMO ambiguity functions.

 \begin{equation*}
\chi^*(\tau,\nu, f'_s, f_s) e^{-j2\pi\nu \tau} = (\sum^{M-1}_{m = 0 } \sum^{M-1}_{{m^{'}} = 0 } \chi(u_m^{'},u_m)(\tau,\nu)  e^{i2\pi \gamma (f^{'}_s m^{'} -f_s m )})^{*}  e^{-j2\pi\nu \tau}
 \end{equation*}
 \begin{equation*}
  \begin{aligned}
= \sum^{M-1}_{m = 0 } \sum^{M-1}_{{m^{'}} = 0 } \chi^*(u_m^{'},u_m)(\tau,\nu)   e^{-j2\pi\nu \tau} e^{i2\pi \gamma (f_s m - f^{'}_s m^{'} )} 
\end{aligned}
\end{equation*}
 \begin{equation*}
  \begin{aligned}
= \sum^{M-1}_{m = 0 } \sum^{M-1}_{{m^{'}} = 0 } \chi(u_m,u_m^{'})(-\tau,-\nu)    e^{i2\pi \gamma (f_s m - f^{'}_s m^{'} )} 
\end{aligned}
\end{equation*}
 \begin{equation*}
= \chi(-\tau,-\nu, f_s, f'_s) = (\chi(\tau, \nu f_s, f'_s) \circ J) \circ J
\end{equation*}
\end{proof}

This property shows that, application of $J$ on MIMO ambiguity functions gives a delay and Doppler mirror image. Next property shows the LFM effect.

\begin{description}
\item[Property 4.3]
\end{description}

\begin{equation*} 
\chi(u,v)(\tau,\nu) \circ t(-k) = \chi(u,v)(\tau,\nu - k \tau)  e^{-j\pi k \tau^{2}} = \chi^{LFM}(u,v)(\tau,\nu)
\end{equation*}

\begin{proof} 
The first equation can be proved by diract substitution. The LFM effect is,

 \begin{equation*}
\chi^{LFM}(u,v)(\tau,\nu) =\int_{-\infty}^{+\infty} (u(t)e^{j2\pi k \nu t^{2}}) (v(t + \tau)e^{j2\pi k \nu (t+\tau)^{2}})^* e^{j2\pi\nu t} dt
 \end{equation*}
 \begin{equation*}
  \begin{aligned}
\chi(u,v)(\tau,\nu - k \tau)  e^{-j2\pi k \tau^{2}} 
\end{aligned}
\end{equation*}
\end{proof} 

The next property shows the effect of $t(-k)$ on MIMO ambiguity funtions.
\newpage

\begin{description}
\item[Property 4.3']
\end{description}
If we fix $f_s,f'_s$ 

\begin{equation*} 
\chi(\tau,\nu, f_s, f'_s) \circ t(-k) =  \chi( \tau, \nu - k \tau, f_s, f'_s) e^{-j\pi k \tau^{2}} = \chi^{LFM}(\tau,\nu, f_s, f'_s) 
\end{equation*}

\begin{proof} 
We will use property 4.3 in definition of MIMO ambiguity functions,

 \begin{equation*}
\chi^{LFM}(\tau,\nu, f_s, f'_s) = (\sum^{M-1}_{m = 0 } \sum^{M-1}_{{m^{'}} = 0 } \chi^{LFM}(u_m,u_m^{'})(\tau,\nu)  e^{i2\pi \gamma (f_s m -f^{'}_s m^{'}  )})^{*}  e^{-j2\pi\nu \tau}
 \end{equation*}
 \begin{equation*}
  \begin{aligned}
= (\sum^{M-1}_{m = 0 } \sum^{M-1}_{{m^{'}} = 0 } \chi^{LFM}(u_m,u_m^{'})(\tau,\nu- k \tau ) e^{-j\pi k \tau^{2}} e^{i2\pi \gamma (f_s m -f^{'}_s m^{'}  )})^{*}  e^{-j2\pi\nu \tau}
\end{aligned}
\end{equation*}
 \begin{equation*}
  \begin{aligned}
= \chi( \tau, \nu - k \tau, f_s, f'_s) e^{-j\pi k \tau^{2}} =\chi(\tau,\nu, f_s, f'_s) \circ t(-k)
\end{aligned}
\end{equation*}
\end{proof}

LFM signals improve range resolution with the ridge effect. Here, we see that $t(-k)$ acts on the ambiguity function to create an LFM effect. Next property shows the effect of $m(b)$, which scales the waveforms.

\begin{description}
\item[Property 4.4]
\end{description}
Let $\tilde {u(t)} = u(bt)$, $\tilde {v(t)} = v(bt)$ then,

\begin{equation*} 
\chi(u,v)(\tau,\nu) \circ m(b) = \tfrac{1}{b} \chi(u,v)(b\tau,\tfrac{\nu}{b})= \chi(\tilde u, \tilde v)(\tau,\nu)
\end{equation*}

\begin{proof} 
We will use a change of variables,

 \begin{equation*}
\chi(\tilde u, \tilde v)(\tau,\nu)  =\int_{-\infty}^{+\infty} u(bt) v^*(b(t + \tau)) e^{j2\pi\nu t} dt
 \end{equation*}
 \begin{equation*}
  \begin{aligned}
=1/b \int_{-\infty}^{+\infty} u(t') v^*(t' + b\tau) e^{j2\pi \tfrac{\nu}{b} t'} dt'
\end{aligned}
\end{equation*}
Where $t'$ dummy variable is not related with that in proof of property 4.2
 \begin{equation*}
  \begin{aligned}
= \tfrac{1}{b} \chi(u,v)(b\tau,\tfrac{\nu}{b})
\end{aligned}
\end{equation*}
Left hand side is by direct substitution.
\end{proof}

Next, we will use this relation in MIMO ambiguity functions.

\begin{description}
\item[Property 4.4']
\end{description}
If we fix $f_s,f'_s$ and let $\chi_{\tilde u}(\tau,\nu, f_s, f'_s)$ denote the MIMO ambiguity function with  $ \tilde u_m = u_m(bt) $,

\begin{equation*} 
\chi_{u}(\tau,\nu, f_s, f'_s) \circ m(b) = \tfrac{1}{b} \chi(b\tau,\tfrac{\nu}{b}, f_s, f'_s)  = \chi_{\tilde u}(\tau,\nu, f_s, f'_s)
\end{equation*}

\begin{proof} 
We will use property 4.4 in definition of MIMO ambiguity functions

 \begin{equation*}
\chi_{\tilde u}(\tau,\nu, f_s, f'_s) = \sum^{M-1}_{m = 0 } \sum^{M-1}_{{m^{'}} = 0 } \chi_{\tilde u}(u_m,u_m^{'})(\tau,\nu)  e^{i2\pi \gamma (f_s m - f^{'}_s m^{'} )} 
 \end{equation*}
 \begin{equation*}
  \begin{aligned}
= \sum^{M-1}_{m = 0 } \sum^{M-1}_{{m^{'}} = 0 } \tfrac{1}{b} \chi(u_m,u_m^{'})(b \tau,\tfrac{\nu}{b} )  e^{i2\pi \gamma (f_s m - f^{'}_s m^{'} )} 
\end{aligned}
\end{equation*}
 \begin{equation*}
  \begin{aligned}
= \tfrac{1}{b} \chi(b\tau,\tfrac{\nu}{b}, f_s, f'_s)
\end{aligned}
\end{equation*}
Left hand side is proved by substituting  $\tfrac{1}{b} \chi(u_m,u_m^{'})(b \tau,\tfrac{\nu}{b} )$with $ \chi(u_m,u_m^{'})(\tau,\nu) \circ m(b) $
\end{proof} 

This property shows the effect of $m(b)$ on the MIMO ambiguity functions which is turning waveforms $u_m(t)$ to the scaled ones $u_m(bt)$.

\section{Conclusion}
In this paper  MIMO ambiguity functions are  related with irreducible unitary representations of $H_{\mathbb{R}}$. Some  harmonic analysis relations found for traditional radar in \cite{auslander1985radar}, \cite{miller2002topics}, \cite{moran2001mathematics} are generalized to MIMO case. In section 3, the norm of MIMO ambiguity functions is shown to be directly determined by norm of individual, seperate signals. It was shown that, the MIMO ambiguity functions create and orthonormal basis of $L_2(\mathbb{R}^{2})$, which can be used for time frequency signal processing applications. Moreover, orthogonality of MIMO ambiguity functions arising from two orthonormal sets was given. In addition to these, results from \cite{auslander1985radar} and \cite{miller2002topics}, concerning positive definiteness and an identity between different functions with same ambiguity functions are used in MIMO ambiguity functions. In section 4, some symmetry relations of ambiguity functions are investigated by use of the action of generators of the group $SL(2,\mathbb{R})$. Signals creating such relations and their corresponding ambiguity functions are studied.

\printbibliography

\end{document}